# A Population-Modulated Bibliometric Measure with an Application in the Field of Statistics

J. Panaretos[1] and C.C. Malesios
*Department of Statistics*
*Athens University of Economics and Business*
*76 Patision St, 10434 Athens Greece*

**Abstract**

We use Confirmatory Factor Analysis (CFA) to derive a unifying measure of comparison of scientists based on bibliometric measurements, by utilizing the h-index, some similar h-type indices as well as other common measures of scientific performance. We use a real data example from nine well-known Departments of Statistics (Berkeley, Carnegie Mellon, Chicago, Duke, Harvard, Minnesota, Oxford, Stanford and Washington) to demonstrate our approach and argue that our combined measure results in a better overall evaluation of a researchers' scientific work.

## 1. INTRODUCTION

A simple way of measuring scientific research impact is often based on the number of publications and the number of citations received by a researcher.

However, these numbers alone fail to capture aspects of a scientist's research record, making it difficult to distinguish the truly influential scientists.

The assessment of research performance of scientists based on citation count proposed by *Hirsch (2005)*, has become the favourite single metric for assessing and validating publication/citation outputs of researchers. Following the introduction of the h-index, numerous articles and reports have appeared, either proposing

[1] e-mail for correspondence: jpan@aueb.gr

modifications of the h-index, or examining its properties and its theoretical background. For an extensive and critical review of the h-index and other similar indices see *Panaretos and Malesios (2008).*

In this paper, we use Confirmatory Factor Analysis (CFA) to derive a population modulated measure to assess scientific research output and impact, by combining the h-index with some of the widely used h-type indices proposed in the literature, as well as with other common measures of scientific performance. We demonstrate our method using data on researchers, affiliated with nine well-known Departments of Statistics (eight in the US and one in the UK) that represent a diverse collection of Departments with different strategies.

The paper is divided into seven parts. Section 2 is a short introduction and overview of the h-index and some of the most significant related indices recently proposed in the literature. In Section 3 we present the data, the new methodology proposed and a review on the related literature. In section 4 we provide some summary statistics on the research output of the Departments of Statistics under study. A description of the construction of the unifying measure and results of the statistical analysis are presented in Section 5, while in Section 6 the results of the analysis on the nine Departments of Statistics are presented. In the final section a discussion of the findings is provided followed by a summary and the main conclusions of the paper.

## 2. THE H-INDEX AND SOME OF ITS GENERALIZATIONS/ MODIFICATIONS

The h-index is based on the number of publications of a researcher, along with the associated citations received by those publications. By definition:

*"A scientist has index h if h of his N papers have at least h citations each, and the other (N - h) papers have at most h citations each".*

Among the advantages of this index is its simplicity and ease of calculation. It attempts to reflect high quality work, since it combines both citation impact (citations received) with publication activity (papers published) and is not affected by a single paper (or a few papers) that has many citations.

The h-index is also not sensitive to lowly cited publications, so a simple increase in the number of publications does not improve the h-index.

There are a number of situations however in which the h-index may provide misleading information about a scientist's output. For instance, the lack of sensitivity of the h-index when it comes to highly-cited papers included in the h-core (the papers that received more than h-citations) is a frequently noticed disadvantage of it. Thus, various modifications and generalizations of the h-index have been appearing in the literature starting almost immediately after its introduction. For full details see *Panaretos and Malesios (2008).*

In what follows, we briefly present some of the most significant modifications associated with the h-index, that are employed for the purposes of our analysis.

*The g-index*

The h-index is robust in the sense that it is insensitive to a set of non-cited (or poorly cited) papers and also to one, or relatively few, outstandingly highly cited papers. That is, once a highly cited article is included in the top h papers of the output of a scientist, the actual number of the paper’s citations does not play a role in the h-index, and any increase of the papers’ citations does not alter the h-index of the scientist.

As a remedy, *Egghe (2006a, 2006b, 2006c)* defined the g-index.

> Definition: “*The g-index is the highest number g of articles that together received $g^2$ or more citations*”

This index is increased by a strongly skewed frequency distribution of the citations; that is, the higher the number of the citations at the top range, the higher the g-index.

*The R- and AR-index*

*Jin et al. (2007)* introduced two modifications of the h-index, namely the R- and AR- indices to eliminate some of the disadvantages of the h-index. The R-index measures the h-core's citation intensity, while the AR-index goes one step further and takes into account the age of each publication. This allows these indices to increase or decrease over time.

If we rank the researcher's articles according to the received citations in decreasing order, then the R-index is defined as:

$$R = \sqrt{\sum_{j=1}^{h} C_j}.$$

where $C_j$ denotes the number of citations received by the j-th article. It is clear from the above definition that always $h \leq R$.

The AR-index is defined as:

$$AR = \sqrt{\sum_{j=1}^{h} \frac{C_j}{\alpha_j}}.$$

where $\alpha_j$ denotes the age of the article j. The advantage of the AR-index is that it gradually suppresses the contribution from articles that have stopped receiving new citations.

*The individual index for correcting for co-authorship*

To correct for the presence of many co-authors in a single publication, *Batista et al. (2006)* divided the h-index by the mean number of researchers in the h publications, i.e., $\overline{N} = N^{(T)} / h$, where $N^{(T)}$ is the total number of authors in the considered h papers (multiple occurrences of authors are allowed), and calls the devised index the $h_I$-index. Similarly, *Hirsch (2005)* proposed the normalization of the h-index by a factor that reflects the average number of co-authors.

## 3. METHODS AND DATA COLLECTION

Although the use of single metrics based on bibliometric measurements for the comparison of scientists has steadily gained popularity in recent years, there is an ongoing debate regarding the appropriateness of such "simple" measures of research performance to measure such complex activities. A recent report by the joint Committee on Quantitative Assessment of Research *(Adler et al., 2008)* argues strongly against the use of citation metrics alone as a tool for assessing quality of research, and encourages the use of more complex methods for judging the impact of scientists (e.g. an assessment based on combining both citation metrics as well as other criteria such as memberships on editorial boards, awards, invitations or peer reviews).

Earlier, along these lines, *Egghe (2007)* noticed that "*the reality is that as time passes, it's not going to be possible to measure an author's performance using just one tool. A range of indices is needed that together will produce a highly accurate evaluation of an author's impact*". An empirical verification of the intuitive view expressed by Egghe came recently by *Bollen et al. (2009)*, who - based on the results of a principal component analysis on a total number of 39 existing indicators of scientific impact - claim that scientific impact is a multi-dimensional notion that cannot be effectively reduced to a single indicator.

In what follows we make an effort to combine the h-index with certain commonly used h-type indices already presented in the previous section, along with the total number of articles and the total number of citations of a researcher. We use CFA to derive a Population-Modulated measure of bibliometric research output performance. We use the term "Population-Modulated" (P-M) because its value for an individual scientist depends on the population against which he is being compared.

There is a steadily increasing literature on applications of factor analysis and Principal Component analysis in Scientometrics in the recent years. For example, in a comparative study of some of the most important h-type indices proposed in the literature, *Bornmann et al. (2008a)* perform an Exploratory Factor Analysis (EFA) using as observed variables nine h-type indicators (including the h-index) in an effort to reveal the latent factors causing the latter indices. In addition to the h-index, the g-, $h^{(2)}$-, A-, R-, AR-, $h_w$-, m-quotient and m-indices, all variants of the h-index, have been utilized to reduce dimensionality and identify possible subsets among these indicators that are more correlated to each other.

In a follow-up study, *Bornmann et al. (2008b)* re-ran the aforementioned EFA, adding in the previously described h-type indicators the two standard indicators in scientometrics, i.e., the total number of articles published and the total number of citations received by these articles. As a tool for assessing the research performance of scientists, the authors proposed the use of any pair of indicators from the two factors, i.e. one indicator that is related to the number of papers in the researcher's productive core and one indicator that is related to the impact of the papers in the researcher's productive core.

*Costas and Bordons (2007)*, also implemented exploratory factor analysis to investigate possible associations of the h-index with other measures of scientific research (measures that describe both quality and quantity of the performance of a researcher) such as the total number of articles, total number of citations, number of citations per article, percentage of highly cited papers, the median impact factor (IF), the normalized position of the publication journal (NPJ), the relative citation rate (RCR) and the percentage of papers with an RCR above 1.

In another recent application of data reduction methodology to bibliometric measures, *Hendrix (2008)* analyzed bibliometric data obtained on the faculty of the Association of the American Medical Colleges member schools, covering the period between 1997 and 2007, and a total of 123 researchers.

We employ a CFA model in order to derive a unified Population-Modulated measure of bibliometric performance of scientists that combines the information provided by not only a single bibliometric index but from a series of measures, such as the Hirsch index and the Hirsch-type indices as well as two basic measures of bibliometric performance (total number of articles and total number of citations) of a researcher. We do this, assuming that there exists a latent dimension underlying the measurable dimensions expressed through the selected bibliometric indicators. The manifest variables are the 6 bibliometric indicators of the researcher's performance and as the latent part of the model we assume a unique underlying factor.

To do this, we collected data on the faculty members of nine University Departments of Statistics: Stanford University, the University of California, Berkeley, the University of Minnesota, Harvard University, the University of Oxford, the University of Washington, Carnegie Mellon University, Duke University and the University of Chicago. The data collection period was between April 2008 and July 2008 and refers to the total number of publications and citations of the faculty serving at the time of the data collection in the nine Departments. We collected information on the current faculty from the web pages of the corresponding Departments. Besides the faculty affiliated with a single Statistics Department, some of the researchers have joint appointments whereas others are courtesy professors. Before developing our methodology we present the data collected, the values of some of the existing indices and some of their descriptive characteristics.

## 4. RESEARCH OUTPUT OF NINE OF THE TOP DEPARTMENTS OF STATISTICS

The data we have collected are publications and related citations of the faculty members of the nine Departments and they are available on the freely accessible citation database "Publish or Perish" (*Publish or Perish User's Manual, 2008*). The data collection period was between April 2008 and July 2008.

We have accumulated this information for the 238 faculty members of the nine Departments. Namely, the Statistics Departments of Stanford University: 30 researchers (faculty: 18, joint appointments: 4, courtesy appointments: 3, consulting: 3, emeritus: 2); the University of California, Berkeley: 41 researchers (faculty: 11, joint appointments: 17, adjunct: 4, emeritus: 9); the University of Minnesota: 21 researchers (faculty: 16, emeritus: 5); Harvard University: 20 researchers (faculty: 11, joint appointments: 6, adjunct: 1, emeritus: 2); the University of Oxford: 30 researchers (faculty: 16, joint appointments: 1, adjunct: 10, emeritus: 2, retired: 1); the University of Washington: 26 researchers (faculty: 25, emeritus: 1); Carnegie Mellon University: 25 researchers (faculty: 22, joint appointments: 1, adjunct: 2); Duke University: 20 researchers (faculty: 12, joint appointments: 5, adjunct: 1, emeritus: 2); and the University of Chicago: 25 researchers (faculty: 12, joint appointments: 9, emeritus: 4). Moreover, we have calculated their h-index values as well as three modifications of the h-index, i.e. the g-index, the individual index $h_I$, and the AR-index[2] from the start of their careers up to the point of the collection period. Along with the AR-index, we list the square root of the AR-index (see Tables 1, 2, A1, A2). We chose those four bibliometric indices because the h-index is the most commonly used index for the comparison of scientists based on bibliometric measures while the other three improve on some of the most serious drawbacks of the h-index, such as the robustness of the h-index to highly cited articles, the problem of co-authorship and the age of the articles. We will use the above information to derive our new Population-Modulated measure.

*Some preliminary findings*

Before defining the new measure we present some descriptive measures of interest from the collected data. In the following figure (Figure 1) the averages on the nine Departments of Statistics of the six output indicators are plotted. As we observe, the index with the least variation is the $h_I$-index, while the g-index is exhibiting significant variation among the Departments. This feature of the g-index has been already underlined by *Egghe (2006b)* who noticed that in general the variances of the g-indices will be much higher than the one of the h-indices in a group of authors. On the other hand, the $h_I$-index values in a group of scientists tend to have less variance possibly due to the fact that smaller h-core outputs (fewer articles in the h-core) tend to have smaller average number of authors in comparison to larger h-core outputs that will probably have a larger average number of authors included in them.

[2] Specifically, PoP calculates a variant of the AR-index of Jin et al. (2007) that differs from Jin's definition in that the citations of all papers are summed instead of only the h-core papers of the author.

***Figure 1:*** *Mean values of the 6 indicators on the 9 Departments of Statistics*

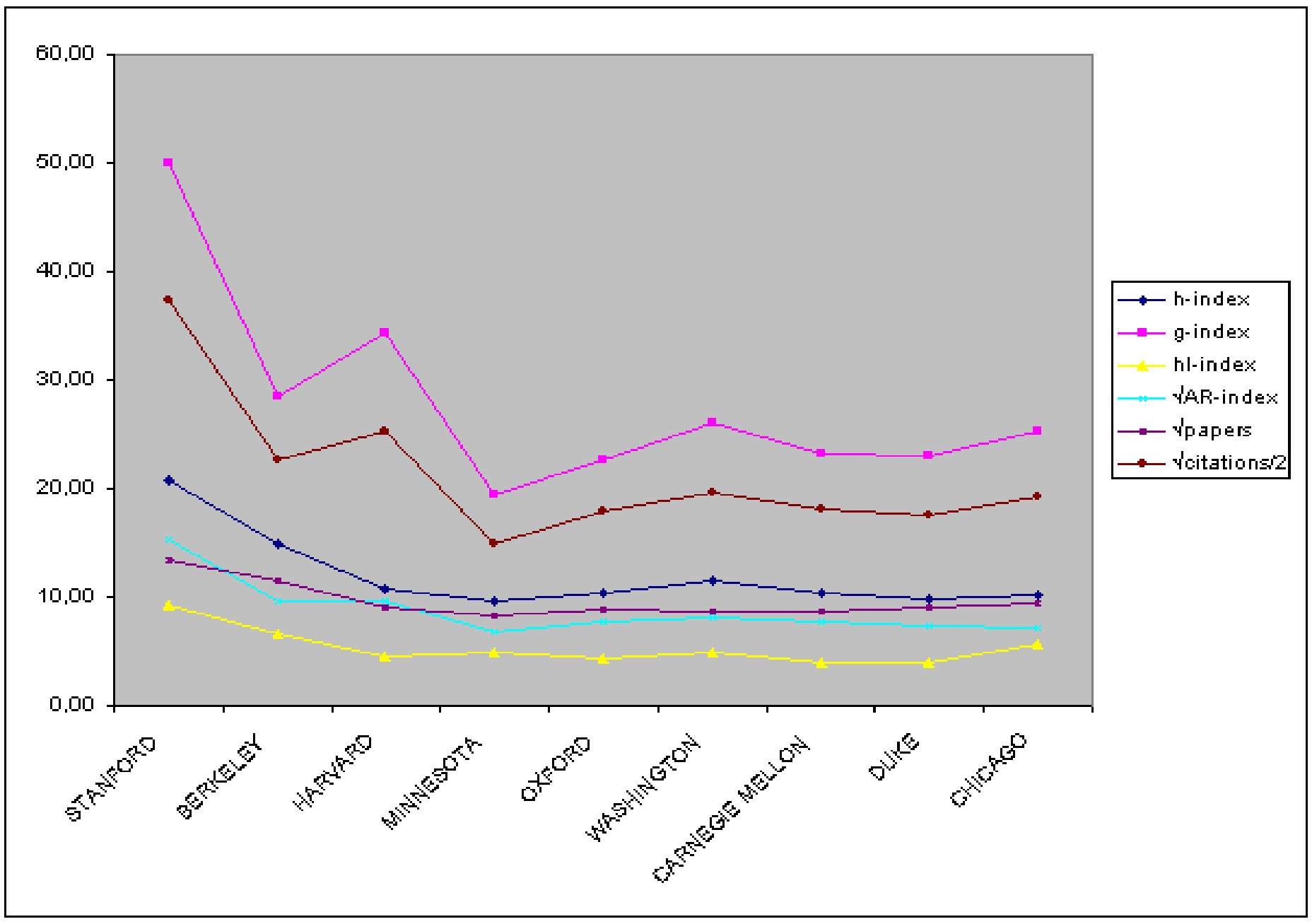

***Table 1:*** *Descriptive statistics of citation metrics for the 9 Departments*

| Department of Statistics | average h-index | median h-index | average g-index | median g-index | average AR-index | median AR-index | average $h_I$-index | median $h_I$-index | average sqrt(AR)-index | median sqrt(AR)-index |
|---|---|---|---|---|---|---|---|---|---|---|
| Stanford | 20.67 | 21.5 | 50 | 52.5 | 356.51 | 121.75 | 9.2 | 8.18 | 15.28 | 11.03 |
| Berkeley | 14.95 | 12 | 28.56 | 26 | 146.78 | 48.54 | 6.56 | 5.79 | 9.61 | 6.97 |
| Harvard | 10.75 | 9.5 | 34.30 | 17.5 | 179.33 | 37.92 | 4.59 | 3.42 | 9.54 | 6.13 |
| Minnesota | 9.71 | 8 | 19.38 | 16 | 93.96 | 28.91 | 4.86 | 3.52 | 6.8 | 5.38 |
| Oxford | 10.37 | 9 | 22.73 | 19.5 | 80.29 | 47.3 | 4.33 | 3.9 | 7.65 | 6.88 |
| Washington | 11.42 | 10.5 | 26.11 | 19 | 100.06 | 67.43 | 4.84 | 4.65 | 8.17 | 8.2 |
| Carnegie Mellon | 10.44 | 8 | 23.28 | 20 | 78.36 | 43.33 | 3.9 | 3.75 | 7.7 | 6.58 |
| Duke | 9.85 | 7 | 23.05 | 13 | 89.72 | 23.22 | 3.89 | 2.5 | 7.45 | 4.8 |
| Chicago | 10.28 | 9 | 25.2 | 19 | 75.62 | 31.86 | 5.65 | 5.4 | 7.18 | 5.64 |
| **Total** | **12.5** | **10** | **28.56** | **20** | **138.35** | **45.69** | **5.51** | **4.27** | **9.03** | **6.76** |

Comparisons between the mean and the median of the indices for the nine Departments of Statistics from Table 1 reveal some interesting information concerning the individual and overall picture of the scientists on each Department as to what concerns the bibliometric indices under study. The large differences between the average and the median index values, appearing in some instances in the table, are due to the fact that the significant majority of scientists have small index values, while only a few scientists have very large index values. Let us examine for instance, the case of Duke University where the 20 researchers of the Department of Statistics have an average h-index of 9.85, while the median h-index is 7. The difference between the two measures is due to the fact that most of the scientists in the Department share relatively small h-index values (14 scientists with h-index less that 11), while only a few scientists have very large h-index values (for instance, A.E. Gelfand, with h-index: 24 or M. West, with h-index: 27). The same holds in the case of the g-index for the faculty of the Department of Statistics of Harvard University, as the large differences between the mean and median g-index value of the Department's staff indicate. While the majority of the 20 researchers (16) have g-indices less than 39, two of the Department's researchers (A.P. Dempster, and D.B. Rubin) have index values 132 and 135, respectively.

As one may observe, the results of citation metrics show an advantage of the Department of Statistics of Stanford University, followed by the Departments of Statistics at Berkeley and Harvard. This is true in all four variables measured, with only a few exceptions, such as those of University of Washington when it comes to average h-index (11.42), or the Universities of Chicago, Minnesota and Washington that outperform the average $h_I$-index of Harvard University.

In general, the Minnesota, Oxford, Washington, Carnegie Mellon, Duke and Chicago Departments of Statistics are very similar in terms of average h-index and average g-index. Also, notice that values of the square root of the AR-index are now comparable with the other three remaining indices.

Our findings cannot be compared with the 1995 ranking of the Statistics Departments by the National Research Council of the USA (*NRC, 1995*). The criteria taken into account in the NRC study were more general. There were 11 criteria (see, e.g. http://www.nap.edu/readingroom/books/researchdoc/summary.html), only one of which was based on publication and citation information of the faculty members of the Departments.

In all, the 238 researchers included in the current study have published (or produced) a total of 29703 papers[3], and have received 386898 citations, until the end of the data collection period (see Tables 2 and A2). In the study period we observe that Stanford is in the first place, when it comes to the average number of papers as well as citations received. In terms of absolute numbers of papers, Berkeley holds the lead.

***Table 2:*** *Descriptive statistics of articles and related citations for the 9 Departments*

| | | papers | | | | | citations | | | | |
|---|---|---|---|---|---|---|---|---|---|---|---|
| **Department of Statistics** | **N** | **average number of papers** | **Std. Deviation** | **median** | **Min** | **Max** | **average number of citations** | **Std. Deviation** | **median** | **Min** | **Max** |
| **Stanford** | 30 | 208.1 | 140.45 | 219.5 | 3 | 581 | 3822.37 | 4106.45 | 2949 | 7 | 17213 |
| **Berkeley** | 41 | 160.76 | 140.59 | 127 | 6 | 553 | 1530.1 | 2664.87 | 849 | 7 | 16516 |
| **Harvard** | 20 | 109.55 | 124.35 | 78.5 | 2 | 474 | 2949.6 | 6360.76 | 362.5 | 4 | 24130 |
| **Minnesota** | 21 | 82.33 | 69.47 | 76 | 5 | 286 | 707.05 | 1250.65 | 276 | 19 | 5522 |
| **Oxford** | 30 | 94.67 | 80.11 | 63.5 | 7 | 300 | 928.2 | 1364.63 | 454 | 13 | 5376 |
| **Washington** | 26 | 91.42 | 78.49 | 67.99 | 3 | 337 | 1230.42 | 1852.01 | 431.5 | 17 | 6518 |
| **Carnegie Mellon** | 25 | 101.52 | 131.65 | 44 | 8 | 635 | 976.8 | 1482.42 | 477,00 | 32 | 5957 |
| **Duke** | 20 | 118.35 | 174.29 | 43.5 | 6 | 741 | 1082.7 | 1630.48 | 239 | 5 | 5395 |
| **Chicago** | 25 | 113.08 | 137.04 | 50.99 | 14 | 647 | 1189.2 | 1916.84 | 369 | 15 | 8566 |
| **Total** | **238** | **124.8** | **128.52** | **79.5** | **3** | **741** | **1625.62** | **3006.34** | **483** | **4** | **24130** |

In an effort to further investigate differences in citations among Departments we collected data on the number of highly cited researchers (HCRs) of the nine Departments in the study. (Information on the HCRs in Statistics affiliated with the Departments in our analysis are available from the Thomson Scientific freely accessible database, covering a 20-year period (1981-1999)). We find that the

[3] The total numbers of publications derived by Publish or Perish include papers published in journals as well as papers that have also been uploaded in web eLibraries (such as arXiv RePEc or SSRN). This may augment the number of publications attributed to a specific researcher. So, there may be duplications that distort the actual publication record of researchers eventually. This observation, to our knowledge, has not been mentioned elsewhere and might be a flaw of the publication/citation databases based on the Google Scholar.

Stanford Statistics Department has a significantly higher number of highly cited statisticians when compared to the other Departments in our study.

Specifically, among the 30 faculty members of the Statistics Department at Stanford, 9 are HCRs, while there are 5 and 4 HCRs in Berkeley and Harvard Statistics, respectively (see Table A3). In addition, among the HCRs of Stanford, 3 HCRs (D.L. Donoho, I.M. Johnstone, IM and R. Tibshirani) are steadily included in the listings of most cited researchers in the field of mathematics, available by the Thomson Scientific for the years 2003 to 2006 (see http://www.in-cites.com/top/index.html; *Ryan and Woodall, 2005*). Table A3 in the appendix presents information on the publication output of the nine Departments. As we can see in this table, the 9 HCRs at Stanford have published 51.9% of the total articles of the Department.

Table A4 presents the results of citation output for the HCRs and the non-HCRs of the nine Departments. As we see, 60.77% of the total number of citations of the faculty of Stanford Statistics is attributed to its 9 HCRs. Similarly the HCRs at Harvard have received 55.99% of the total citations of the faculty of the Department.

In the other 7 Departments the contribution of the HCRs ranges between 2.5% and 38%. The apparent impact of the research work of the HCRs of Stanford is also verifiable by the enormous average number of citations per HCR (7742.5), while, for instance, the corresponding number for Berkeley is 2142.8. One might be tempted to link this to the different hiring policies of the two Departments. Stanford usually hires well-established scientists at the senior level while Berkeley opts for younger promising ones.

The number of the HCRs of Stanford, and their publications and citations is influencing significantly its standings in the tables. Another interesting observation is that the non-HCRs of the Department of Statistics, UC Berkeley publish more frequently in comparison to the non-HCRs of Stanford.

## 5. A POPULATION-MODULATED MEASURE FOR RESEARCH OUTPUT

In the sequel, we choose the 6 variables (i.e. the total number of articles, the total number of citations, the h-index and the related g-, $h_I$- and √AR-indices) as research indicators of the researchers of the nine Departments of Statistics, and we assume that they are observed outcomes of an underlying indicator, which we call Population-Modulated measure (P-M measure). We use this name because it tuned to the peculiarities of the particular scientific population under study.

For practical reasons, of comparisons between the h and h-type indices and the standard bibliometric measures (number of publications and number of citations), we don't use the raw data on the numbers of publications and numbers of citations. Instead, we use the square root of the number of publications and the square root of number of citations divided by 2, respectively. This is justified by the fact that the h-index is proportional to the square root of the number of publications (see *Glänzel, 2006*), while it has been found empirically that the number of citations is proportional to the square of the h-index (see *Hirsch, 2005; van Raan, 2006*). In other words $N_{citations}=\alpha h^2$ approximately, where α is a constant (According to Hirsch the relation between the h-index and the number of citations is dependent on the form of the particular distribution of the data under study, and he empirically noticed that the constant α for the discipline of Physics ranges between 3 and 5, while *van Raan (2006)*, employing regression analysis, found the approximate relation $h= 0.42\times N^{0.45}_{citations}$, using data from the field of Chemistry). For our citation data on mathematics we use α=2 as a suitable constant of proportionality. Our choice is based on the result of a simple regression analysis which suggested the following approximate relation between the h-index and $N_{citations}$: $h \approx 0.4 \times \sqrt{N}$. Other data on highly cited mathematicians found in the literature, provide similar results (using data from *Iglesias and Pecharromán (2007)* on Spanish highly cited mathematicians we obtained the following relation: $h=0.6\times\sqrt{N}$). Furthermore, 2*g/3 scores were used as input data in place of the original g-index values, following *Rousseau (2006)* who has established the latter relationship between the h-index and the g-index for theoretical models, in case where the g-index values are relatively large.

In order to derive the Population-Modulated measure, we assume a one-factor CFA model that at the scientist level can be expressed as:

$$x_{ij}=\lambda_j \xi+ \delta_{ij} \quad (i=1,2,\ldots,238;\, j=1,2,\ldots,6)$$

where $x_{ij}$ denotes the jth bibliometric index of the ith scientist, $\xi$ is the (1×1) scalar of the (unknown) single common factor, the $\lambda_j$'s terms are the factor loadings to be estimated connecting $\xi$ to the the $x_{ij}$'s, and $\delta_{ij}$ is the measurement error in $x_{ij}$ (i.e. the part of $x_{ij}$ that cannot be accounted by the underlying factor $\xi$). It is further assumed that the error terms $\delta_{ij}$ and the common factor $\xi$ have a zero mean and that the common and unique factors are uncorrelated, i.e. $E(\xi-E\xi)(\delta_{ij}-E\delta_{ij})=0$, for every i,j. In vector notation, to scientist i corresponds a (6×1) vector of bibliometric indices $\mathbf{X}_i$:

$$\mathbf{X}_i = \mathbf{\Lambda}\xi + \boldsymbol{\delta}_i$$

where $\mathbf{\Lambda}=(\lambda_1, \lambda_2,...,\lambda_6)^t$, and $\boldsymbol{\delta}=(\delta_{i1}, \delta_{i2},..., \delta_{i6})^t$.

In turn, the CFA model based on the complete data set can be written as:

$$\mathbf{X}=\mathbf{\Xi}\mathbf{\Lambda}^t+\mathbf{\Delta}$$

where $\mathbf{X}$ is the (238×6) matrix of bibliometric indices for the 238 scientists, $\mathbf{\Xi}=\xi\mathbf{1}$, where boldface $\mathbf{1}$ is a (238×1) vector of 1's, $\mathbf{\Lambda}^t$ is the transpose of the (6×1) vector of factor loadings, and finally, $\mathbf{\Delta}$ denotes the (238×6) matrix of measurement errors.

Then, the $(238^2\times6^2)$ variance-covariance matrix of the data denoted by $\mathbf{\Sigma}$ is given by:

$$\begin{aligned}
\boldsymbol{\Sigma} &= Cov\left(\boldsymbol{\Xi}\boldsymbol{\Lambda}^t + \boldsymbol{\Delta}\right) = Cov\left(\xi\mathbf{1}\boldsymbol{\Lambda}^t + \boldsymbol{\Delta}\right) = \\
&= Cov\left(\xi\mathbf{1}\boldsymbol{\Lambda}^t\right) + Cov\left(\boldsymbol{\Delta}\right) = \\
&= E\left[\left(\xi\mathbf{1}\boldsymbol{\Lambda}^t - E\left(\xi\mathbf{1}\boldsymbol{\Lambda}^t\right)\right)\otimes\left(\xi\mathbf{1}\boldsymbol{\Lambda}^t - E\left(\xi\mathbf{1}\boldsymbol{\Lambda}^t\right)\right)\right] + E\left[\left(\boldsymbol{\Delta} - E\left(\boldsymbol{\Delta}\right)\right)\otimes\left(\boldsymbol{\Delta} - E\left(\boldsymbol{\Delta}\right)\right)\right] = \\
&= E\left[\left(\xi - E\left(\xi\right)\right)\mathbf{1}\boldsymbol{\Lambda}^t \otimes \left(\xi - E\left(\xi\right)\right)\mathbf{1}\boldsymbol{\Lambda}^t\right] + E\left[\boldsymbol{\Delta}\otimes\boldsymbol{\Delta}\right] = \\
&= E\left[\left(\xi - E\left(\xi\right)\right)^2 \cdot \left(\mathbf{1}\boldsymbol{\Lambda}^t \otimes \mathbf{1}\boldsymbol{\Lambda}^t\right)\right] + \boldsymbol{\Theta} = \\
&= \left(\mathbf{1}\boldsymbol{\Lambda}^t \otimes \mathbf{1}\boldsymbol{\Lambda}^t\right) E\left[\left(\xi - E\left(\xi\right)\right)^2\right] + \boldsymbol{\Theta} = \\
&= \left(\mathbf{1}\boldsymbol{\Lambda}^t \otimes \mathbf{1}\boldsymbol{\Lambda}^t\right)\sigma^2 + \boldsymbol{\Theta},
\end{aligned}$$

where $\sigma^2$ denotes the $(1\times1)$ scalar of the variance of single factor $\xi$, and $\boldsymbol{\Theta}$ denotes the $(238^2\times6^2)$ variance-covariance matrix of the measurement errors. It is assumed by the CFA model that $E(\boldsymbol{\Delta}^t\boldsymbol{\Xi}) = \mathbf{0}$, $E(\boldsymbol{\Xi}^t\boldsymbol{\Delta}) = \mathbf{0}$, $E(\boldsymbol{\Delta}^t\boldsymbol{\Delta}) = \boldsymbol{\Theta}$, $E(\boldsymbol{\Xi}^t\boldsymbol{\Xi}) = \Phi$ and $E(\mathbf{X}^t\mathbf{X}) = \boldsymbol{\Sigma}$. The aim is to estimate the unknown elements of $\boldsymbol{\Lambda}$, $\sigma^2$ and $\boldsymbol{\Theta}$.

Such a model is usually fit by maximum likelihood[4]. If we denote by $\mathbf{S}$ the empirical covariance matrix of the matrix of the observed variables $\mathbf{X}$ (i.e. the sample variance-covariance matrix), then to obtain the ML estimates of $\boldsymbol{\Lambda}$, $\sigma^2$ and $\boldsymbol{\Theta}$, one needs to maximize the following likelihood function:

$$\log\mathrm{L}(\boldsymbol{\Lambda},\sigma^2,\boldsymbol{\Theta}) = -\frac{1}{2}n\left[\log\left|\boldsymbol{\Sigma}\right| + tr\left(\mathbf{S}\boldsymbol{\Sigma}^{-1}\right)\right].$$

In fact, it has been shown (*Jöreskog, 1969*) that maximizing logL is equivalent to minimizing the following function:

$$F\left(\boldsymbol{\Lambda},\boldsymbol{\Phi},\boldsymbol{\Theta}\right) = \log\left|\boldsymbol{\Sigma}\right| + tr\left(\mathbf{S}\boldsymbol{\Sigma}^{-1}\right) - \log\left|\mathbf{S}\right| - 6,$$

where 6 is the order of the vector $\boldsymbol{\Theta}$.

[4] The six bibliometric indicators used for the fit of the CFA model are subject to moderate or high non-normality. However, a series of studies based on simulated data have shown that ML estimation can produce valid results even under non-normal conditions (*Jöreskog, 1990; Benson and Fleishman, 1994*).

Maximum Likelihood Estimation for our CFA model is carried out using the LISREL 8.8 (*Jöreskog and Sorbom, 1999*) software. The derived estimated CFA model is displayed via the path diagram presented in Figure 2.

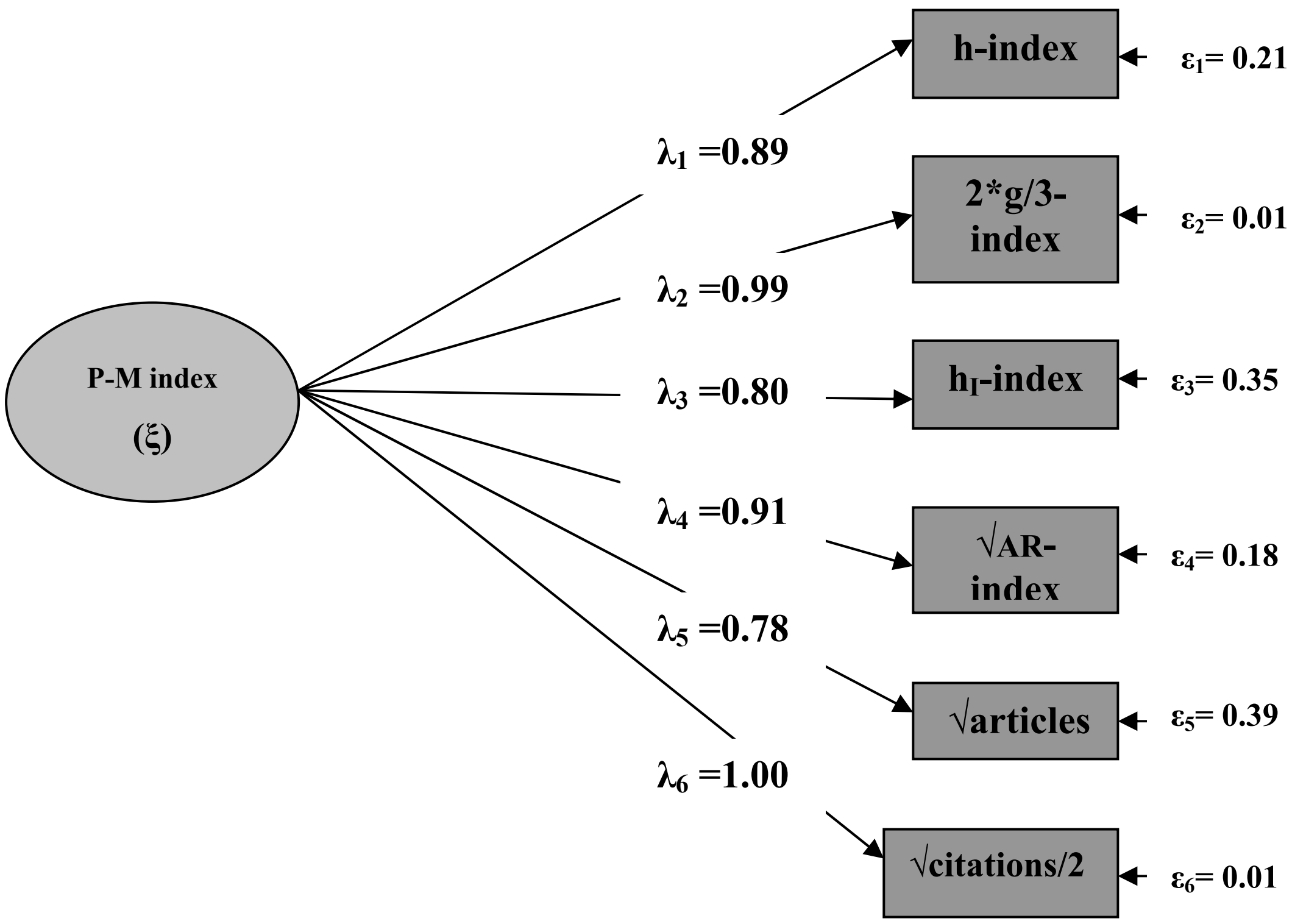

**Figure 2**: path diagram of the estimated CFA model.
[Squares represent the observed variables (h-index, 2*g/3-index, $h_I$-index, square root of AR-index, square root of number of articles, square root of number of citations/2). The ellipse represents the latent factor].

The above (one-factor) model is displayed as a diagram in which squares represent the observed variables and the ellipse represents the latent variable. Single-headed arrows are used to imply a direction of assumed causal influence between the latent and the observed variables. Numerical values along each arrow correspond to the (standardised) factor loadings of each observed variable on the latent variable.

By observing the path diagram of the model and Table 3 several things are immediately apparent. Firstly, all factor loadings are significant at the 5% significance level (all unstandardised loadings are at least twice the size of the standard errors of the estimates). Clearly, a large proportion of the variance in each observed variable is accounted for by the fitted model as the $R^2$ values of Table 3 reveal ($R^2$ denotes the percentage of each one of the manifest variables explained by the CFA model). Accordingly, all six observed variables are related to the latent factor. Not all variables however are equally related to the P-M measure. Higher loadings are observed on the √citations/2 and the g-index (1.00 and 0.99, respectively), whereas the lowest loading is observed on the "total number of articles" manifest variable (0.78).

***Table 3:*** *Summary statistics of the CFA model fit*

| Manifest variables | Unstandardized loadings | Standard error | p-value | $R^2$ |
|---|---|---|---|---|
| **h-index** | 8.15 | 0.46 | <0.05 | 0.79 |
| **2*g/3-index** | 16.89 | 0.79 | <0.05 | 0.99 |
| **hI-index** | 3.47 | 0.23 | <0.05 | 0.65 |
| **√AR-index** | 6.85 | 0.38 | <0.05 | 0.82 |
| **√articles** | 4.07 | 0.28 | <0.05 | 0.61 |
| **√citations/2** | 18.43 | 0.84 | <0.05 | 0.99 |

With regards to the model's adequacy, the fit indices values for the evaluation of the goodness-of-fit of the model obtained from LISREL indicate that the factor model tested provides a moderate fit to the 6 observed variables: GFI[5]=0.69, NNFI=0.71, NFI=0.83, CFI=0.83 (accepted boundaries for an excellent fit >0.90).

*The P-M measure as a combination of the h-index and related indices*

As already stated, factor analysis models have received a lot of attention both in theory and in practice. Within this framework, it has become common practice to estimate individual factor scores (*Bartholomew and Knott, 1999*) and utilize them for subsequent analyses.

[5] GFI: Goodness of fit index, NNFI: Non-Normed fit index, NFI: Normed fit index, CFI: Comparative fit index

For instance, factor scores for the latent variables can be first predicted and then used as variables in ANOVA and OLS regression (as dependent or explanatory variables) (e.g. *Urban and Hauser, 1980*) or as input data to cluster (e.g. *Funkhouser, 1983*) and discriminant analysis (*Horton, 1979*).

The CFA model of the analysis fulfils the requirement of strong associations between observed and latent variables (with only one exception, where factor loadings appear slightly lower). LISREL was used to derive factor scores (*Mels, 2004*) of the first-order CFA model, on the 6 observed variables data. There are several methods available for estimating latent variable scores (see, e.g. *Bartholomew and Knott, 1999*). LISREL uses the procedure of *Anderson and Rubin (1956)* as described in *Jöreskog (2000)* for estimating latent variable scores. This procedure has the advantage of producing latent variable scores that have the same covariance matrix as the latent variables themselves. However, because the mean of the factor is set to zero, it is natural that some estimated factor scores are negative and some positive.

To avoid deriving negative factor scores (and subsequently use these negative scores as scientific performance indices), we chose to calculate the component scores instead of the factor scores. These were introduced by *Bartholomew and Knott (1999)* and are defined as follows. For any single latent variable and a collection of manifest variables, say $X_i$ (i=1,2,…,p), we can produce estimates of the underlying latent variable by using a linear combination of the responses, known as the component scores, which are given by:

$$y = \sum_{i=1}^{p} \lambda_i x_i ,$$

where $\lambda_i$ denotes the factor's loading on the i-th manifest variable, and the $x_i$ is the observed value of the i-th manifest variable (i=1,2,…,p).

The procedure for the calculation of the Population-Modulated (P-M) measure of each scientist is as follows: from the fitting of the CFA model, the unstandardised factor loadings (which are common to all scientists) are derived for the 6 indicators on the underlying factor. These are subsequently utilized as weights indicating the relative importance of the different bibliometric variables to take into account.

The P-M measure for say, scientist i, is essentially the linear combination of the six bibliometric variables scores of the specific scientist, using as coefficients the factor loadings corresponding to each manifest variable.

(This linear combination of the individual scores on the manifest variables with the overall factor loadings is usually known as the component score).

Thus, the P-M measure is not identical to the values of the underlying factor that gives the factor analysis, but instead uses the obtained loadings of the manifest indicators on the hypothesized underlying factor as a way to construct a proxy for the unobservable $\xi$.

For each of our 6 manifest variables, the estimated (unstandardised) factor loadings were: for the h-index 8.15 (0.46), for the 2g/3-index 16.89 (0.79), for the √AR-index 6.85 (0.38), for the $h_I$-index 3.47 (0.23), for the √articles 4.07 (0.28) and for the √citations/2 18.43 (0.84). The values in the parentheses are the corresponding standard errors. Thus, the calculation of the P-M index for, say the ith researcher (i=1,2,….,238) in our study, will be based on the following equation:

$$(P-M)_i = \frac{8.15}{0.46}\times\left(h-index\right)_i + \frac{16.89}{0.79}\times\left(\tfrac{2}{3}\,g-index\right)_i + \frac{6.85}{0.38}\times\left(\sqrt{AR}-index\right)_i + \\ + \frac{3.47}{0.23}\times\left(h_I-index\right)_i + \frac{4.07}{0.28}\times\left(\sqrt{articles}\right)_i + \frac{18.43}{0.84}\times\left(\sqrt{citations}/2\right)_i .$$

At this point we have to note that there is not an absolute P-M measure for an individual scientist. The measure changes depending on the group we are studying, hence the name Population-Modulated measure.

The advantage is that the measure requires working with a well-defined population and prevents comparisons between people in disparate scientific fields. To do something like this we need to pool together the two (or more) populations, which will reveal the differences between the two fields.

It also allows for bibliometric comparisons between two different disciplines. It is a data-dependent way to see how seriously we should consider one index compared to another, for the different fields of research.

The P-M measure can also help in finding whether a particular index is better suited for a particular discipline as opposed to another. In this way, we can quantify statements like “the h-index is not suitable for mathematics”. It would also be of interest to compute the P-M measure for those who have joint appointments in two different Departments (e.g., Statistics and Electrical Engineering and Computer science or Statistics and Biostatistics) and see how the P-M measure of these scientists changes from one group to another.

For example, let us say that we have a series of three bibliometric indices x, y and z. Then, if we derive the following linear combinations of the three indices (P-M measure) for two groups of scientists, one say belonging to Statistics and the other to Bibliometrics:

$$\begin{aligned} &ax + \beta y + \gamma z \qquad (Statistics) \\ &a'x + \beta' y + \gamma' z \qquad (Bibliometrics) \end{aligned}$$

and compare coefficients $\alpha$ with $\alpha'$, $\beta$ with $\beta'$ and $\gamma$ with $\gamma'$, if $\alpha >> \alpha'$ then x is taken more heavily into account in statistics than in bibliometrics.

## 6. AN APPLICATION OF THE NEW BIBLIOMETRIC MEASURE IN THE FIELD OF STATISTICS

In terms of the above, individual component scores (i.e. the P-M scores) were derived for the 238 researchers of the nine Statistics Departments in our study. For comparisons between the 238 researchers we have divided the component scores by 100. In this way, the combined P-M measure scores are more comparable to the h-index values, as well as to the other three h-type indices used in our analysis.

In the following table (Table 4), descriptive statistics of the derived scores are presented for the 9 Departments. We observe that Harvard and Stanford have the highest standard deviations. This is again attributed to the presence of a number of highly cited researchers. Two of the Department's researchers at Harvard (D.B. Rubin and A.P. Dempster) score quite high in the P-M measure (ranked 2nd and 5th respectively), widening in this way the gap in the P-M measures of the Department's researchers. This explains the difference between the average and the median P-M measure values. At Sanford also we have a significant number of highly cited researchers (9).

**Table 4:** *Average component scores for the 9 Departments*

| | | P-M Measure | | | | |
|---|---|---|---|---|---|---|
| **Department of Statistics** | **N** | **mean** | **Std. Deviation** | **median** | **Min** | **Max** |
| **Stanford** | 30 | 25.06 | 15.02 | 26.12 | 1.71 | 62.85 |
| **Berkeley** | 41 | 16.06 | 11.05 | 15.19 | 1.38 | 62.38 |
| **Oxford** | 30 | 12.32 | 7.68 | 10.99 | 2.21 | 33.5 |
| **Washington** | 26 | 13.53 | 9.51 | 10.81 | 2.13 | 37.96 |
| **Carnegie Mellon** | 25 | 12.4 | 8.07 | 10.7 | 3.25 | 34.67 |
| **Chicago** | 25 | 13.15 | 9.16 | 10.14 | 2.71 | 36.43 |
| **Harvard** | 20 | 16.07 | 16.56 | 9.86 | 1.23 | 62.57 |
| **Minnesota** | 21 | 10.92 | 8.21 | 9.14 | 2.55 | 37.76 |
| **Duke** | 20 | 12.13 | 10.02 | 7.73 | 1.38 | 33.8 |
| **Total** | **238** | **14.97** | **11.54** | **11.32** | **1.23** | **62.85** |

A small section of the output obtained by calculating the P-M measure for individual scientists is presented in the following table (Table 5) for comparisons between the already calculated Hirsch type indices and the newly proposed P-M measure. (The number in the parenthesis indicates the ranking for the particular index).

***Table 5:*** *Faculty members ranked by component scores (P-M measure) derived by the CFA model*

| University | Author | h-index | 2*g/3-index | √AR-index | $h_I$-index | √papers | √citations/2 | P-M measure |
|---|---|---|---|---|---|---|---|---|
| Stanford | Donoho, D.L. | 52 (2) | 86 (3) | 41.64 (2) | 25.75 (1) | 24.1 (3) | 92.77 (3) | 62.85 (1) |
| Harvard | Rubin, D.B. | 29 (10) | 103.33 (1) | 39.79 (3) | 12.94 (15) | 14.53 (33) | 109.84 (1) | 62.57 (2) |
| Berkeley | Jordan, M.I. | 60 (1) | 82.67 (4) | 41.85 (1) | 21.05 (3) | 23.52 (4) | 90.87 (4) | 62.38 (3) |
| Stanford | Tibshirani, R. | 40 (3) | 75.33 (5) | 37.83 (4) | 15.53 (7) | 17.29 (19) | 80.61 (5) | 52.56 (4) |
| Harvard | Dempster, A.P. | 21 (17) | 88 (2) | 23.21 (12) | 10.02 (29) | 12.57 (42) | 93.61 (2) | 50.6 (5) |
| Stanford | Stork, D.G. | 20 (18) | 66.67 (6) | 32.51 (7) | 7.27 (38) | 15.97 (24) | 71.3 (6) | 42.72 (6) |
| Stanford | Friedman, J.H. | 31 (8) | 61.33 (7) | 22.92 (13) | 11.31 (20) | 19.1 (12) | 67.11 (7) | 41.94 (7) |
| Stanford | Diaconis, P. | 39 (4) | 51.33 (9) | 20.78 (16) | 19.5 (4) | 20.32 (7) | 58.41 (9) | 40.34 (8) |
| Washington | Raftery, A.E. | 32 (7) | 52.67 (8) | 24.93 (10) | 12.19 (18) | 14.97 (30) | 57.09 (10) | 37.96 (9) |
| Minnesota | Cook, R.D. | 29 (10) | 47.33 (13) | 34.45 (6) | 15.29 (8) | 16.91 (20) | 52.55 (16) | 37.76 (10) |
| Stanford | Wong, W-H. | 30 (9) | 50 (11) | 26.16 (9) | 9.09 (32) | 22.11 (6) | 56.22 (11) | 37.64 (11) |
| Stanford | Hastie, T.J. | 25 (13) | 48 (12) | 36.47 (5) | 8.45 (35) | 17.41 (17) | 52.35 (17) | 36.56 (12) |
| Chicago | Billingsley, P. | 15 (23) | 61.33 (7) | 14.37 (36) | 14.06 (11) | 10.95 (47) | 65.44 (8) | 36.43 (13) |
| Stanford | Efron, B. | 28 (11) | 50 (11) | 19.77 (19) | 15.68 (6) | 16.25 (22) | 54.5 (14) | 35.9 (14) |
| Harvard | Liu, J.S. | 29 (10) | 48 (12) | 23.77 (11) | 10.13 (26) | 19.44 (9) | 53.4 (15) | 35.76 (15) |
| Berkeley | Aldous, D.J. | 34 (5) | 41.33 (16) | 16.9 (33) | 24.6 (2) | 19.31 (10) | 47.96 (24) | 34.95 (16) |
| Carnegie Mellon | Lehoczky, J.P. | 28 (11) | 50.67 (10) | 18.08 (26) | 10.05 (28) | 14.66 (32) | 54.58 (13) | 34.67 (17) |
| Duke | West, M. | 27 (12) | 42 (15) | 20.72 (17) | 9.23 (30) | 27.22 (1) | 49.93 (23) | 33.8 (18) |
| Oxford | Cox, D.R. | 32 (7) | 40 (18) | 19.29 (21) | 13.84 (13) | 17.32 (18) | 51.02 (20) | 33.5 (19) |
| Duke | Gelfand, A.E. | 24 (14) | 48 (12) | 19.59 (20) | 9 (33) | 14.18 (34) | 51.94 (18) | 32.86 (20) |
| Washington | Stuetzle, W. | 20 (18) | 51.33 (9) | 19.95 (18) | 4.6 (47) | 13.49 (38) | 54.92 (12) | 32.82 (21) |
| Washington | Bookstein, F.L. | 22 (16) | 47.33 (13) | 17.54 (29) | 10.08 (27) | 13.93 (36) | 51.01 (21) | 31.92 (22) |
| Carnegie Mellon | Fienberg, S. | 23 (15) | 46 (14) | 17.14 (31) | 9.12 (31) | 15.65 (26) | 50.48 (22) | 31.73 (23) |
| Oxford | Lauritzen, S. | 21 (17) | 48 (12) | 17.09 (32) | 9 (33) | 12.57 (42) | 51.85 (19) | 31.62 (24) |
| Stanford | Chui, C.K. | 29 (10) | 38.67 (20) | 16.81 (34) | 14.75 (10) | 15.87 (25) | 44.32 (28) | 30.69 (25) |
| Stanford | Lavori, P. | 33 (6) | 40.67 (17) | 19.1 (22) | 7.07 (41) | 11.92 (44) | 44.49 (27) | 30.54 (26) |
| Chicago | Goodman, L.A. | 25 (13) | 38.67 (20) | 12.32 (44) | 13.89 (12) | 25.44 (2) | 44.68 (26) | 30.51 (27) |
| Stanford | Anderson, T.W. | 25 (13) | 41.33 (16) | 10.24 (49) | 17.86 (5) | 14.7 (31) | 45.52 (25) | 29.93 (28) |
| Stanford | Johnstone, I.M. | 22 (16) | 38 (21) | 29.63 (8) | 8.2 (37) | 14.53 (33) | 41.75 (34) | 29.87 (29) |
| Stanford | Dembo, A. | 25 (13) | 39.33 (19) | 18.05 (27) | 10.96 (22) | 15.56 (27) | 43.88 (29) | 29.63 (30) |
| Berkeley | Pitman, J.W. | 30 (9) | 32 (24) | 17.67 (28) | 15 (9) | 19.92 (8) | 41.51 (35) | 29.61 (31) |
| Stanford | Cover, T.M. | 28 (11) | 38.67 (20) | 12.98 (42) | 12.85 (16) | 15.03 (29) | 42.7 (32) | 29.06 (32) |
| Berkeley | Speed, T.P. | 21 (17) | 40 (18) | 22.14 (15) | 7.11 (40) | 14 (35) | 43.49 (30) | 28.91 (33) |
| Chicago | Niyogi, P. | 24 (14) | 37.33 (22) | 22.43 (14) | 8.73 (34) | 13.34 (39) | 40.72 (36) | 28.47 (34) |
| Stanford | Olkin, I. | 23 (15) | 39.33 (19) | 11.12 (48) | 10.37 (24) | 15.26 (28) | 43.36 (31) | 27.79 (35) |
| Berkeley | Peres, Y. | 28 (11) | 28 (29) | 17.19 (30) | 10.59 (23) | 19.16 (11) | 37.16 (41) | 26.58 (36) |
| Berkeley | Bartlett, P.L. | 25 (13) | 31.33 (25) | 18.47 (24) | 8.45 (35) | 18.08 (15) | 36.64 (42) | 26.4 (37) |
| Chicago | Stephens, M. | 14 (24) | 38 (21) | 13.49 (41) | 6.76 (42) | 18.38 (13) | 41.8 (33) | 25.9 (38) |
| Berkeley | Stone, C.J. | 24 (14) | 35.33 (23) | 11.92 (46) | 8.23 (36) | 12.69 (41) | 38.9 (38) | 25.58 (39) |
| Berkeley | Rice, J. | 18 (20) | 32 (24) | 14.62 (35) | 5.79 (46) | 23.15 (5) | 38.84 (39) | 25.43 (40) |
| Oxford | Silverman, B. | 25 (13) | 31.33 (25) | 13.56 (40) | 10.59 (23) | 16.64 (21) | 35.6 (43) | 25.4 (41) |
| Duke | Sacks, J. | 21 (17) | 35.33 (23) | 12.77 (43) | 7.23 (39) | 13.6 (37) | 38.9 (38) | 25.18 (42) |
| Harvard | Chernoff, H. | 18 (20) | 37.33 (22) | 8.34 (50) | 12 (19) | 11.62 (45) | 40.28 (37) | 25.01 (43) |
| Stanford | Lai, T-L. | 25 (13) | 26 (31) | 12.27 (45) | 13.59 (14) | 17.75 (16) | 34.73 (44) | 24.45 (44) |
| Oxford | Snijders, T. | 14 (24) | 35.33 (23) | 18.2 (25) | 6.13 (45) | 9.33 (48) | 37.83 (40) | 23.9 (45) |
| Minnesota | Geyer, C. | 17 (21) | 30.67 (26) | 13.91 (38) | 10.32 (25) | 13.11 (40) | 34.07 (45) | 23.02 (46) |
| Berkeley | Dudoit, S. | 18 (20) | 30 (27) | 18.72 (23) | 6.23 (44) | 11.27 (46) | 33.11 (46) | 22.82 (47) |
| Washington | Richardson, T. | 19 (19) | 28 (29) | 14.15 (37) | 6.33 (43) | 18.36 (14) | 33.08 (47) | 22.78 (48) |
| Duke | Clark, J.S. | 16 (22) | 28.67 (28) | 13.63 (39) | 11.13 (21) | 16.09 (23) | 33.04 (48) | 22.69 (49) |
| Minnesota | Hawkins, D.M. | 24 (14) | 26.67 (30) | 11.88 (47) | 12.26 (17) | 12.41 (43) | 30.87 (49) | 22.52 (50) |

The authors appearing in the output have been ranked according to their P-M indices, derived by the CFA model (the top 50 authors). All other 6 bibliometric measures are included in the table for comparisons.

As is easily seen, ranking of bibliometric performance of scientists according to the P-M measure is slightly different when compared to the h-index ranking. Let us take for example, D. Donoho of Stanford University and M. Jordan of the University of California, Berkeley. Donoho has an h-index of 52, while Jordan has an h-index of 60. However, when it comes to P-M measure bibliometric comparison, D. Donoho is ranked slightly higher (P-M measure: 62.85 compared to a 62.38 of M. Jordan).

Indeed, if we observe the number of articles published and citations received we see that D. Donoho with 581 articles (√articles=24.1) has 17213 citations (√citations/2=92.77), while M. Jordan with 553 articles (√articles=23.52) has 16516 citations (√citations/2=90.87). Further, as concerns the g-index, which is a measure of the h-core's citation intensity, the values are 127 for D. Donoho and 124 for M.

Jordan. Also, when comparing the $h_I$-index which is a measure of normalization for the co-authorship, D. Donoho has $h_I$=25.75 and M. Jordan 21.05. Finally, as concerns the √AR-index, the two have roughly equal values (41.85, compared to 41.64).

Another example is provided by the output research comparison between F. Bookstein, of University of Washington and I. Johnstone, of Stanford University. Bookstein has a P-M measure value of 31.92 while Johnstone's is 29.87. F. Bookstein has a lead in three out of six metrics (the g-index, the $h_I$-index and the number of citations) while I. Johnstone has higher √AR-index and a higher total number of published papers. Both share the same h-index. Overall, L. Booksteins's h-core papers are more cited, he has less recent work, and has published fewer papers with fewer co-authors that have received a higher number of citations.

Thus, one may argue that rating bibliometric performance according to the new combined index results in a better overall evaluation of the researchers' work, since it takes into account not just one index (such as the h-index), or any other single bibliometric measure (e.g. total number of citations), but is a combination of the main indices that measure different aspects of the overall work of the researcher.

## 7. CONCLUSIONS AND DISCUSSION

An important consideration in evaluating research performance of a scientist is the multiple manifestations of his/her work. So, as many authors have argued, the use of indices to assess only a component of a scientist's work (e.g. citations) is unfair to scientists (see, e.g. *Adler et al., 2008*; *Kelly and Jennions, 2006; Sanderson, 2008*). Measuring the research performance of a scientist by using only his/hers bibliometric data is already more or less restrictive by default, let alone by measuring the citation performance with only a single one of the metrics already described.

Our results indicate that the method of rating scientific performance based on bibliometric measurements introduced in this paper enhances the index ranking based on measuring a single characteristic. Our measure provides a more general picture of the scientist's activity, by combining h-type indices proposed in the literature with older bibliometric measures, such as the total number of articles and total number of citations.

Moreover, the new measure provides some additional discriminatory power for research output comparisons, and we argue that ranking according to the P-M measure is perhaps more fair, when compared to the ranking of scientists based on each one of the single bibliometric/citation metrics (new and old ones), separately.

Of course, this measure is cumbersome to calculate. In addition, there does not exist an absolute and single P-M measure for an individual scientist, since the measure can change, depending on the specific population of researchers selected for its calculation. The magnitude of the value of the P-M measure of a scientist is relative to the P-M measures of the other scientists under study. It does allow though one to produce a ranking for a given set of scientists irrespective of their fields.

However, given that single indices fail to capture important aspects of research performance, measures such as the P-M are useful when comparisons of scientists based on bibliometric parameters alone are necessary.

Besides the obvious use of the P-M measure as a tool for the assessment of the overall performance of a researcher, it can also be utilized for interdisciplinary comparisons; by analyzing two (or more) sets of researchers of different fields of research, we can derive useful information associated with the magnitude of effectiveness of each one of the bibliometric indices on the specific scientific field and provide insight about the behavior of citation data in each discipline using the calculated composite measure. In this way, the strong and weak performance of the single bibliometric indicators on each scientific field of research can be identified and we can quantify statements, until now only intuitively imposed, such as “the *x*-index is (or is not) suitable for the *y* discipline”, or “the *x*-index is accounted more largely in the *y* discipline than in the *z* discipline”.

The P-M measure could also produce valid results when implemented for the comparisons between different sub-fields of the same discipline (i.e. comparisons of the P-M measure and the single indicators performance in Biostatistics, applied Statistics and Statistical theory).

The choice of the specific indicators to be included in the calculation of the P-M measure is of significant importance too, and will be the subject of future research. Revising some of the already included indicators for calculation of the P-M measure or expanding the list of the six indicators with new single metrics could enhance the accuracy of the performance of the proposed measure by covering other aspects of the researcher’s work, not depicted by the already considered indicators.

# APPENDIX

**Table A1:** *Descriptive statistics of citation metrics for the 9 Departments*

| | | h-index | | | | | g-index | | | | | hI-index | | | | | AR-index | | | | |
|---|---|---|---|---|---|---|---|---|---|---|---|---|---|---|---|---|---|---|---|---|---|
| **Department of Statistics** | **N** | **mean** | **Std. Deviation** | **median** | **Min** | **Max** | **mean** | **Std. Deviation** | **median** | **Min** | **Max** | **mean** | **Std. Deviation** | **median** | **Min** | **Max** | **mean** | **Std. Deviation** | **median** | **Min** | **Max** |
| **Stanford** | 30 | 20.67 | 12.16 | 21.5 | 2 | 52 | 50 | 32.71 | 52.5 | 2 | 129 | 9.2 | 6 | 8.18 | 1 | 25.75 | 356.51 | 470.99 | 121.75 | 0.44 | 1733.49 |
| **Berkeley** | 41 | 14.95 | 10.74 | 12 | 2 | 60 | 28.56 | 21.77 | 26 | 1 | 124 | 6.56 | 5.04 | 5.79 | 0.33 | 24.6 | 146.78 | 282.65 | 48.54 | 0.14 | 1751.05 |
| **Harvard** | 20 | 10.75 | 8.56 | 9.5 | 1 | 29 | 34.3 | 41.99 | 17.5 | 2 | 155 | 4.59 | 3.97 | 3.42 | 0.5 | 12.94 | 179.33 | 367.78 | 37.92 | 1 | 1583.07 |
| **Minnesota** | 21 | 9.71 | 6.9 | 8 | 2 | 29 | 19.38 | 15.93 | 16 | 3 | 71 | 4.86 | 3.73 | 3.52 | 1.5 | 15.29 | 93.96 | 254.82 | 28.91 | 1.91 | 1186.81 |
| **Oxford (UK)** | 30 | 10.37 | 6.61 | 9 | 2 | 32 | 22.73 | 16.45 | 19.5 | 3 | 72 | 4.33 | 2.86 | 3.9 | 1 | 13.84 | 80.29 | 97.61 | 47.3 | 1.07 | 372.08 |
| **Washington** | 26 | 11.42 | 7.19 | 10.5 | 2 | 32 | 26.11 | 21.75 | 19 | 3 | 79 | 4.84 | 2.78 | 4.65 | 0.4 | 12.19 | 100.06 | 142.95 | 67.43 | 2.33 | 621.36 |
| **Carnegie Mellon** | 25 | 10.44 | 6.76 | 8 | 3 | 28 | 23.28 | 17.63 | 20,00 | 4 | 76 | 3.9 | 2.6 | 3.75 | 1 | 10.05 | 78.36 | 87.45 | 43.33 | 2.59 | 326.8 |
| **Duke** | 20 | 9.85 | 7.67 | 7 | 1 | 27 | 23.05 | 21.12 | 13 | 2 | 72 | 3.89 | 3.2 | 2.5 | 0.44 | 11.13 | 89.72 | 125.9 | 23.22 | 0.12 | 429.45 |
| **Chicago** | 25 | 10.28 | 6.32 | 9 | 2 | 25 | 25.2 | 21.49 | 19 | 2 | 92 | 5.65 | 3.74 | 5.4 | 0.75 | 14.06 | 75.62 | 109.47 | 31.86 | 3.01 | 503.26 |
| **Total** | **238** | **12.79** | **8.79** | **11** | **1** | **60** | **27.6** | **20.73** | **22** | **1** | **155** | **5.66** | **4.15** | **4.66** | **0.33** | **25.75** | **115.34** | **200.41** | **46.47** | **0.12** | **1751.05** |

**Table A2:** *Descriptive statistics of articles, square root of articles, citations and square root of citations divided by 2 for the 9 Departments*

| | | papers | | citations | | √papers | | √citations/2 | |
|---|---|---|---|---|---|---|---|---|---|
| Department of Statistics | N | total number of papers | average number of papers | total number of citations | average number of citations | total number of √papers | average number of √papers | total number of √citations/2 | average number of √citations/2 |
| **Stanford** | 30 | 6243 | 208.1 | 114671 | 3822.37 | 401.1 | 13.37 | 1119.62 | 37.32 |
| **Berkeley** | 41 | 6591 | 160.76 | 62734 | 1530.1 | 470.63 | 11.48 | 924.56 | 22.55 |
| **Harvard** | 20 | 2191 | 109.55 | 58992 | 2949.6 | 180.24 | 9.01 | 506.08 | 25.3 |
| **Minnesota** | 21 | 1729 | 82.33 | 14848 | 707.05 | 173.47 | 8.26 | 314.03 | 14.95 |
| **Oxford** | 30 | 2840 | 94.67 | 27846 | 928.2 | 267.2 | 8.9 | 535.46 | 17.85 |
| **Washington** | 26 | 2377 | 91.42 | 32003 | 1230.88 | 227.49 | 8.75 | 511.15 | 19.66 |
| **Carnegie Mellon** | 25 | 2538 | 101.52 | 24420 | 976.8 | 219.01 | 8.76 | 454.31 | 18.17 |
| **Duke** | 20 | 2367 | 118.35 | 21654 | 1082.7 | 179.42 | 8.97 | 352.47 | 17.62 |
| **Chicago** | 25 | 2827 | 113.08 | 29730 | 1189.2 | 234.49 | 9.38 | 481.52 | 19.26 |
| **Total** | **238** | **29703** | **124.8** | **386898** | **1625.62** | **2353.06** | **9.89** | **5199.2** | **21.84** |

**Table A3:** *Results on the published articles output of the 9 Departments of statistics*

| Department of Statistics | Number of faculty | Total number of articles | Average number of articles | Number of non-HCRs | Total number of articles (non-HCRs) | Average number of articles (non-HCRs) | percentage of total number of articles (non-HCRs) | Number of HCRs | Total number of articles (HCRs) | Average number of articles (HCRs) | percentage of total number of articles (HCRs) |
|---|---|---|---|---|---|---|---|---|---|---|---|
| **Stanford** | 30 | 6243 | 208.1 | 21 | 3003 | 143 | 48.1% | 9 | 3240 | 360 | 51.9% |
| **Berkeley** | 41 | 6591 | 160.76 | 36 | 5509 | 153.03 | 83.58% | 5 | 1082 | 216.4 | 16.42% |
| **Harvard** | 20 | 2191 | 109.55 | 16 | 1047 | 65.44 | 47.8% | 4 | 1144 | 286 | 52.21% |
| **Minnesota** | 21 | 1729 | 82.33 | 20 | 1443 | 72.15 | 83.46% | 1 | 286 | 286 | 16.54% |
| **Oxford** | 30 | 2840 | 94.67 | 28 | 2382 | 85.07 | 83.87% | 2 | 458 | 229 | 16.13% |
| **Washington** | 26 | 2377 | 91.42 | 24 | 2064 | 86 | 86.83% | 2 | 313 | 156.5 | 13.17% |
| **Carnegie Mellon** | 25 | 2538 | 101.52 | 24 | 2448 | 102 | 96.45% | 1 | 90 | 90 | 3.55% |
| **Duke** | 20 | 2367 | 118.35 | 18 | 1997 | 110.94 | 84.37% | 2 | 370 | 185 | 15.63% |
| **Chicago** | 25 | 2827 | 113.08 | 24 | 2673 | 111.38 | 94.55% | 1 | 154 | 154 | 5.45% |
| **Total** | **238** | **29703** | **124.8** | **211** | **22566** | **106.95** | **75.97%** | **27** | **7137** | **264.33** | **24.03%** |

**Table A4:** *Results on the citations output of the 9 Departments of statistics*

| Department of Statistics | Number of faculty | Total number of citations | Average number of citations | Number of non-HCRs | Total number of citations (non-HCRs) | Average number of citations (non-HCRs) | percentage of total number of citations (non-HCRs) | Number of HCRs | Total number of citations (HCRs) | Average number of citations (HCRs) | percentage of total number of citations (HCRs) |
|---|---|---|---|---|---|---|---|---|---|---|---|
| **Stanford** | 30 | 114671 | 3822.37 | 21 | 44988 | 2142.29 | 39.23% | 9 | 69683 | 7742.56 | 60.77% |
| **Berkeley** | 41 | 62734 | 1530.1 | 36 | 52020 | 1445,00 | 82.92% | 5 | 10714 | 2142.8 | 17.08% |
| **Harvard** | 20 | 58992 | 2949.6 | 16 | 25963 | 1622.69 | 44.01% | 4 | 33029 | 8257.25 | 55.99% |
| **Minnesota** | 21 | 14848 | 707.05 | 20 | 9326 | 466.3 | 62.81% | 1 | 5522 | 5522 | 37.19% |
| **Oxford** | 30 | 27846 | 928.2 | 28 | 17263 | 616.54 | 61.99% | 2 | 10583 | 5291.5 | 38.01% |
| **Washington** | 26 | 32003 | 1230.88 | 24 | 22857 | 952.38 | 71.42% | 2 | 9146 | 4573 | 28.58% |
| **Carnegie Mellon** | 25 | 24420 | 976.8 | 24 | 22409 | 933.71 | 91.76% | 1 | 2011 | 2011 | 8.24% |
| **Duke** | 20 | 21654 | 1082.7 | 18 | 14827 | 823.72 | 68.47% | 2 | 6827 | 3413.5 | 31.53% |
| **Chicago** | 25 | 29730 | 1189.2 | 24 | 28976 | 1207.33 | 97.46% | 1 | 754 | 754 | 2.54% |
| **Total** | **238** | **386898** | **1625.62** | **211** | **238629** | **1130.94** | **61.68%** | **27** | **148269** | **5491.44** | **38.32%** |